\documentstyle[epsfig,aps]{revtex}
\begin{document}
\title{ Extending Linear Response: Inferences from Electron-Ion Structure Factors}
\author{A.A. Louis and N.W. Ashcroft}
\address{ Cornell Center for Materials Research,
and Laboratory of Atomic and Solid State Physics,
Cornell University, Ithaca, NY 14853-2501}
\date{\today}
\maketitle
\begin{abstract}
Linear response methods applied to electron systems often display a level of
accuracy which is notable when viewed in terms of the strengths of perturbing
interactions.  Neglect of higher response terms is in fact justifiable in
 many cases and it can be shown to stem from an intrinsic interference between
  atomic and electronic length
scales.  For fluid metallic systems it can be further shown that electron-ion
structure (increasingly accessible experimentally) can be understood from an
application of {\em linear response} in the electron system, combined with
hard-sphere
like correlation for the ionic component.

\vspace{4pt}
\end{abstract}
\vspace{20pt}


The nearly-free-electron (NFE) approximation underlies much of our understanding
of the properties of condensed matter, in particular simple metals.  While
ab-initio simulation techniques have long superseded the NFE approximation in
quantitative accuracy, it remains an important source of insight and of
simplifying concepts to elucidate qualitative trends across different materials.
It also provides guidance in situations that remain out of the reach of
computational ab-initio techniques\cite{Hafn92}.  For many years, the density
$\rho^{ind}({\bf \vec k})$ of an initially uniform electron gas induced by an embedded
pseudo-potential $v^{ps}({\bf \vec k})$ has been successfully treated at linear order even
though $v^{ps}({\bf \vec k})$ is not necessarily a small perturbation.  The linear approach
is a key component in many applications of the NFE approximation, examples of
which include pseudo-potential calculations of the free-energy of simple metals,
their relative structural stability (and corresponding cohesive properties) and
also the determination of effective ion-ion
potentials\cite{Hein70,Ashc78,Hafn83,Hafn87}.  The accuracy of the latter is a
particularly striking example of the efficacy of linear response; while the
energy scale of unscreened ions at typical separations is of the order of Ry, linear screening leads to 
ion-ion potentials fully capable of describing observed structural phase
transitions and implying consequent energy scales of the order of mRy.

Here we address the evident success of the linear approximation which to date remains incompletely resolved. We show that
the implied neglect of higher order response is supported by physical
arguments.  In particular we explicitly demonstrate that the nonlinear terms
are small for specific cases, and give arguments to suggest that this may 
be expected to hold more generally, the main exception being hydrogen.
As an application of the underlying argument, but one with experimental consequences, we demonstrate that simple
linear-response theory augmented by a hard-sphere approximation for ionic
structure leads to a quantitatively accurate analytical representation of
electron-ion structure factors $S_{ei}(k)$ in liquid metals, these now in
principle accessible through recent advances in both
neutron and x-ray scattering techniques.  Another route to effective
electron-ion interactions therefore opens, but here through the fluid state.

To begin, consider the response of the interacting electron gas to a single ion,
where the electron-ion interaction is modeled by a {\em pseudo-potential},
taken as a simple local one-parameter empty-core form\cite{Ashc66} i.e.:
$v^{ps}(k) = -(4 \pi e^2/k^2) cos(kR_c)$, where the {\em atomic} core-radius
$R_c$ or equivalently the zero-crossing $k_0 = \pi/2R_c$ is typically fixed by
an atomic property such as the ionization energy (or by a measurable crystalline metallic
property such as the Fermi surface\cite{Pseudo}).  The pseudo-potential leads to
a local electron density inhomogeneity representable by $\rho^{ind}(k)$.  There
are two routes to represent this induced density, the first (esentially exact) from solving the
familiar  self-consistent Kohn-Sham equations within the
local density approximation (LDA) \cite{Kohn65}, and the second from the standard expansion of the
response in powers of the perturbing (pseudo)potential, i.e.;
\begin{equation}\label{eq1}
 \rho^{ind}(k) = \chi_1(k) v^{ps}(k) +
\sum_{\vec{k}_1} \sum_{\vec{k}_2} \chi_2(k,k_1,k_2)v^{ps}(k_1)v^{ps}(k_2)
+....
\end{equation} 
Here the response functions $\chi_n(k_1...)$ are properties of the {\em
homogeneous} interacting electron gas, the first being the well known linear
response function\cite{Hafn87}.  The second is given by:  \vspace{.5cm}
\begin{equation}\label{eq2} 
\chi_2(k_1,k_2,k_3) = \left(\chi_2^0(k_1,k_2,k_3) +
\frac{1}{2} \mu_2(k_1,k_2,k_3)\chi_1^0(k_1)\chi_1^0(k_2)\chi_1^0(k_3) \right)
/\left( \epsilon(k_1) \epsilon(k_2) \epsilon(k_3) \right),
\end{equation}
\noindent
where  $\epsilon(k)$ is the usual dielectric function:
 $\epsilon(k) = 1 - \left(4 \pi e^2/k^2 + \mu_1(k)\right)\chi_1^0(k)$. 
In (2) the $\chi_n^0(k_1...)$ are the non-interacting response functions 
(known to second order\cite{Lloy68}) and the $\mu_n(k_1..k_{i+1})$ 
are the homogeneous limits of the $n$-th  functional derivatives of the
exchange-correlation potential  with respect to density.
In particular,
 $\mu_1(k)$ is related
to the spin-averaged local field correction (LFC),
 $G(k) = (k^2/4\pi e^2) \mu_1(k)$.
Fig. 1 compares the full LDA response, and equation (\ref{eq1}), taken to second order; note that it appears to capture most of the complete response with considerable accuracy.

\begin{figure}
\begin{center}
\epsfig{figure=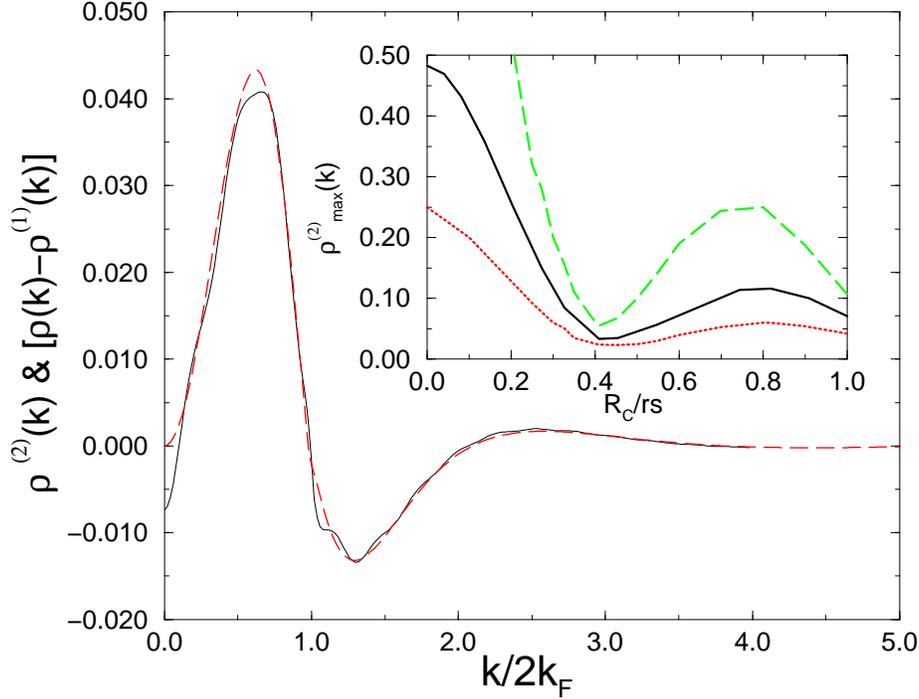,width=12cm}
\vglue0.1cm
\caption{
A comparison of full non-linear LDA response $[\rho(k)$$-$$\rho^{(1)}(k)]$
(solid line) to second order LDA response (dashed line) for an empty core
pseudo-potential with $R_c$$=$$1.5a_0$ embedded in an electron gas with density
parameter $r_s$$=$$3a_0$.  For the scale, compare this to the full response with
the limit $\rho(k$$=$$0)$$=$$1$.  The higher order response is of
the order of a few $\%$ of the full response and in turn,
the second order response
captures almost all the non-linear response.  
(The small difference at $k\rightarrow 0$ is a numerical artifact stemming from the use of a large but finite real-space cut-off radius utilized in the Kohn-Sham procedure.)
In the insert is plotted the maximum of the 2nd order response vs.
$R_c/r_s$ for $r_s$$=$$2a_0$(dotted), $r_s$$=$$3a_0$(solid) and $r_s$$=$$5a_0$(dashed).
Note especially the minimum at $R_c/r_s$$=$$0.41$ which corresponds to
$k_0$$=$$2k_F$.  It is reduced by an order of magnitude from the value
at $R_c$$=$$0$ (hydrogen) and is traced to an interference between atomic
and electronic length-scales.
}
\label{fig4.1.5}
\end{center}
\end{figure}
\vglue-.5cm

Interestingly enough, the combined effects of exchange and correlation partially
cancel between first and second order, and this implies that the neglect of the
$\mu_n(k_1..k_{i+1})$ at {\em both} orders (the RPA) is found to be more accurate
than the result obtained by merely including them at a single order
only\cite{thesis}.  This has important implications for the widespread
application of linear response theory in the derivation of effective ion-ion
potentials in (simple) metals; {\em the neglect of higher order response results
in an overestimate of the role of exchange and correlation}.  The accuracy of
the second order response depicted in Fig.  1 also implies that the use of more
accurate LFC's could, in some cases, lead to an improvement in accuracy over a
full Kohn-Sham LDA calculation.  Although much effort has gone into obtaining
LFC's beyond the ($k=0$) LDA limit at linear order\cite{LFC}, the second order
LFC, directly related to $\mu_2(k_1,k_2,k_3)$, to date remains unknown beyond
the LDA form.  However, the second order electron LFC is the direct analog of
the third order direct correlation function $c^{(3)}(k_1,k_2,k_3)$ of classical
liquid-state theory for which various successful approximations based on lower
order correlations functions have been derived\cite{Akhe97}.  (Since the electron
liquids are more weakly correlated than their classical
counterparts\cite{Ashc95}, it might now be suggested that application of these
classically inspired approaches to the electronic case would be useful.)

A central question now arises (whose answer is important to the proposition we make on electron-ion structure):  Why is the {\em non-linear} response
contribution depicted in Fig.  1 evidently so small?  An
immediate possibility is that higher order terms in equation~(\ref{eq1}) are
large, but actually vary in sign and therefore mutually cancel, order by
order.  But another is that the higher order terms are each {\em individually}
very small.  The success at the level of 2nd order response evidently implies
that the latter is the case:  {\em we find that the response series converges
very rapidly.}  This might be physically anticipated since a larger
atomic-parameter $R_c$ implies a smaller perturbing potential, and the
non-linear response shows a clear decline with increasing $R_c$.  
As anticipated the second order response is found to be largest for $R_c=0$ (hydrogen), but as
$R_c$ increases from zero a noticable secondary minimum occurs when the inverse
atomic length $k_0$ is equal to $2k_F$.  For the cases plotted in Fig.  1, the
secondary minimum is reduced by an entire {\em order of magnitude} when compared
with the value calculated for hydrogen, and is typically factor of three
lower than the secondary maximum at larger $R_c$.  This minimum is attributed to
the following; the second order response function, $\chi_2(k,k_1,k_2)$, itself peaks
when the summed arguments in~(\ref{eq1}) are close to $2k_F$\cite{thesis}.  Accordingly, if
the pseudo-potential zero-crossing $k_0$ is near the response peaks at $2k_F$, a
maximal cancellation or {\em maximal destructive interference of the atomic and
electronic lengthscales} occurs, leading to a minimum in second order response.
We may now postulate that for the simple metals a similar interference effect occurs for the higher order
terms of~(\ref{eq1}).

Typically the value of $k_0/2k_F$ lies between
$0.75$ and $1$, and is therefore very close to the secondary minimum in the
non-linear response.  Note that the ratio of the atomic and electronic length
scales is set primarily by the volume energy terms in the total ground state
energy, and is almost independent of structure\cite{Hafn87}.  
This clarifies in large part why the ubiquitous linear response approximation
performs so remarkably well for many materials and why the higher order terms
are indeed small.  The NFE approximation has often been justified in a context
far wider than linear response alone by appeal to the fact that for a
crystalline solid, the structure dependent reciprocal lattice vectors are
typically near the pseudo-potential zero-crossing $k_0$ with the inference that
the net scattering is small\cite{Hafn92}.  This important effect stems from the
confluence an {\em atomic} and a {\em structural} length scale; the interference
effect we discuss is complementary, but has a different physical origin, namely
an interference between intrinsic {\em atomic} and {\em electronic}
lengthscales.  Once again, the clear exception is
the singular case of a point-charge $(v^{ps}(k) \sim 4 \pi e^2/k^2)$, i.e.  the
case of hydrogen, which has no well-defined core-length scale $k_0$, no
oscillations in the potential and thus no interference effect in the higher
order terms.  In sharp contrast to other systems, non-linear response terms are
large {term by term}.  In fact, the response series may not even formally
converge and great care must be taken when applying concepts derived from
linear-response theory to hydrogen (it is not a simple material).

As noted, the continued accuracy of linear response is important to an interpretation now proposed for electron-ion structure factors
$S_{eI}(k)$ in metallic fluids, these being defined as $k$-space density-density
correlation functions\cite{Ashc78}.  Invoking the adiabatic approximation
they can always be rewritten in terms of the ion-ion structure factors as
follows:
\begin{equation}\label{eq5}
S_{eI}(k) = \frac{n(k)}{\sqrt{Z}} S_{II}(k),
\end{equation} 
which defines a new dimensionless object $n(k)$.  Electron-ion correlations can
therefore be described by convolving the pseudo-electron density (or
pseudo-atom) $n(k)$ with the ionic correlations.  The accuracy of linear
response for the pseudo-potential in an electron-gas implies that it should
now also be an excellent approximation for a determination of the pseudo-atom density.  For simple
liquid metals $S_{II}(k)$ is very well approximated by the Percus-Yevick
analytic form for hard spheres by specifying a single parameter, the packing
fraction $\eta$, which is close to $\eta \sim 0.46$ for most simple metals near
melting\cite{Ashc66b}.  Using this in~(\ref{eq5}), we compare our approach in Fig. 2 to
the {\em full ab-initio} Car-Parrinello\cite{Car85} calculations of de Wijs {\em
et al}\cite{deWi95}.  The correspondence is striking, especially when we note that the parameters
$\eta$ and $R_c$ are {\em a priori} set by other physical properties (no fitting
is necessary).

\begin{figure}
\begin{center}
\epsfig{figure=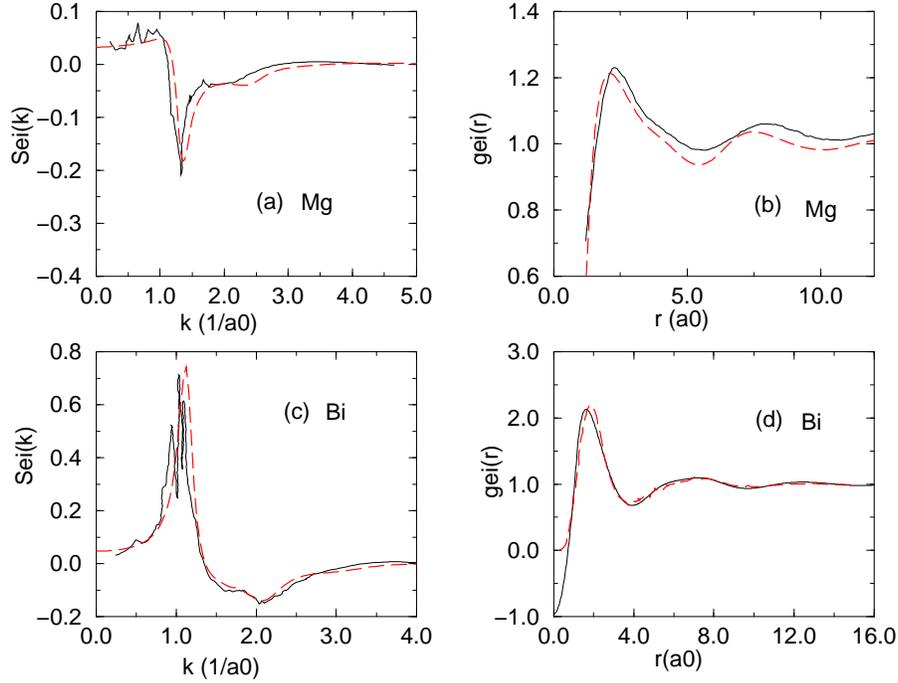,width=12cm}
\vglue0.1cm
\caption{The electron-ion structure factor $S_{eI} (k)$ and related electron-ion
correlation function $g_{ei}(r)$ for Mg and Bi:  Car-Parrinello results of de
Wijs {\em et al}{\protect\cite{deWi95}} (solid line) vs.  the simple
linear-response approach augmented by a hard-sphere approximation (dashed line).
Panel (a) shows $S_{eI}(k)$ and panel (b) shows $g_{eI}(r)$ for liquid Mg.
Panel (c) shows $S_{eI}(k)$ and panel (d) shows $g_{eI}(r)$ for liquid Bi.  For
Mg the parameters (taken from the literature) are:  $r_s$$=$$2.66a_0$,
$R_c$$=$$1.31a_0$ and for Bi the parameters (taken from the literature) are:
$r_s$$=$$2.25a_0$ and $R_c$$=$$1.15a_0$.  Both have a packing fraction
$\eta$$=$$0.46$ (note that for the $g_{eI}(r)$ the region inside the core
radius is not physically significant).
}
 \label{figfullBi_Mg}
\end{center}
\end{figure}
\vglue-.5cm

Besides a semi-quantitative description of electron-ion structure factors, this
linear response theory now provides an important qualitative insight into
the form of the electron-ion structure factors\cite{Loui97b}.  The pseudo-atom
density $n(k)$ is typically largest for small $k$ and rapidly declines for
larger $k$, while the near classical ion-ion structure factor $S_{II}(k)$
follows an inverse behavior; it is small for small $k$.  Together with the
product form~(\ref{eq5}) this
implies that the shape of the electron-ion structure factor $S_{eI}(k)$ is determined
primarily by the the position of the {\em zero-crossing} $\bar{k}_0$ of $n(k)$
with respect to the {\em first maximum} $k_p$ of $S_{II}(k)$.  If $\bar{k}_0 <
k_p$, then $S_{II}(k)$ selects (or filters) the negative part of $n(k)$ and
$S_{eI}(k)$ takes a form similar to that of Mg (Fig.  2 
(a)).  Conversely, if $\bar{k}_0 > k_p$, then the ion-ion structure factor
selects (or filters) the positive part of $n(k)$, and again, $S_{eI}(k)$ takes a
form similar to that of Bi (Fig.  2 
~(\ref{figfullBi_Mg}
 (c)). Since
$\chi_1(k)$ is positive definite, the zero-crossing in linear response occurs at
$k_0$.  The large slope of $n(k)$ near the zero-crossing then implies
that non-linear corrections {\em must} have a small effect on the location
of the zero-crossing, and together with the expected accuracy of linear response
this implies that $\bar{k}_0 \sim k_0$.  As mentioned earlier, for most metals,
$k_0$ is just a little less than $2k_F$, and the latter's ratio to $k_p$ is well
known:  for small valence $(Z \leq 2)$, $2k_F < k_p$; for large valence $(Z\geq
3)$:  $2k_F > k_p$\cite{Zima72}.  This accounts in a straightforward way for the
two separate forms found by deWijs {\em et al}\cite{deWi95}:  For Mg, $\bar{k}_0
< k_p$ $(Z=2)$, which belongs to the {\bf low-valence class} of electron-ion
structure factors.  For Bi, $\bar{k}_0 > k_p$ $(Z=5)$ and we may refer to this
as the {\bf high valence class} of electron-ion structure
factors\cite{deWijssei}.  Generally ions of valence $Z \leq 2$ belong to the low
valence class while ions with valence $Z > 3 $ belong to the high valence class.
Ions with valence $Z=3$ typically belong to the high valence class also,
although they may characterized by a crossover form\cite{thesis}.  The
analytical approach above can easily be extended by using the modern theory
of liquids to obtain improved ion-ion structure factors\cite{Cusa76}, but to
include second order contributions to the pseudo-atom $n(k)$ necessitates not
only second order electron response, but also contributions from ion-ion triplet
structure.  This can also be carried out with concepts from the theory of
classical liquids.\cite{thesis}

These observations have a potentially useful experimental consequence:  the
principal features of electron-ion structure factors can be measured by
exploiting the differences between x-ray scattering, which probes the
density fluctuations of {\em all} electrons, and neutron scattering which generally
probes fluctuations of the nuclei.\cite{Egel74}.  X-ray measurements are 
usually interpreted using a free atom form factor, while our analysis 
suggests that for liquid metals, they should be interpreted with the 
pseudoatom as a form factor.  When this is taken into account, a small
difference between x-ray and neutron scattering determinations of the
ion-ion structure factor should emerge.  This difference is largest 
for metals with a high ratio of valence to core electrons.  For Li (1:2)
or Al (3:10), we predict a 2\% difference at the first peak of the structure
factor\cite{Anta98}, but the largest effects are expected for Be which has
the highest ratio of valence to core electrons (1:1) and for which the 
difference could be as much as 7\%, well within experimental range.  In
addition, Be may straddle the two classes ($k_0$ is near $k_p$), which 
means that small differences in $\bar{k}_0$ with respect to $k_0$ may
lead to significant, qualitative differences in $S_{ei}(k)$, making it 
a particularly interesting candidate for illuminating nonlinear effects.
In a similar way we now anticipate that higher order effects can be 
revealed in partially covalent liquid metals, silicon and gallium 
being examples. The arguments presented suggest that these should become
relatively less important upon an increase in density (via pressure).

The arguments and associated analyis above therefore provide a physical basis for
understanding why linear response theories in dense electron systems generally perform so well.  
The accuracy of linear response is
demonstrated for fluid metals by a simple analytical linear-response theory
augmented by a hard-sphere approach to classical electron-ion structure factors,
which already gives semi-quantitative accuracy.  It suggests that there are two
main classes of electron-ion correlation functions, one for high and one for low
valence metals.  Finally is it suggested that experimental advances in x-ray and
neutron-scattering may be poised to provide
measurements of these electron-ion correlation functions, and hence on the
interactions themselves.

This work was supported by the NSF under Grant No.  DMR96-24-8330.  We
especially thank Professor K. Jacobsen for making an LDA Kohn-Sham program available
to us, and Professor J.-P. Hansen and Dr. D. Muller for useful comments.


\begin{thebibliography}{99}
\bibitem{Pseudo}
  This one-parameter form greatly facilitates our investigations and while it
fails to reproduce the exact scattering properties of the full ion with core
electrons (exemplified by non-local ab-intio pseudo
potentials), we are mainly interested in the response of an electron-gas to an
embedded ion and not in the details of the ions themselves; this slight loss of
accuracy does not weaken the physics.  Note that that the general zero-crossing
$k_0$ of a local $v^{ps}(k)$ is quite well established physically, it is
similar for different representations of $v^{ps}(k)$, and it is not an artifact of
the particular form we choose.
\end{thebibliography}
\end{document}